\documentclass{elsart}
\usepackage{amssymb}

\usepackage{amsfonts}
\usepackage{amsmath}

\input{tcilatex}
\begin{document}

\begin{frontmatter}%

\title{Counting Statistics of Parallel Aluminum-atom Wires}%

\author{Ying-Tsan Tang and Yu-Chang Chen}%

\address{Department of E lectrophysics, National Chiao Tung University, 1001 Ta Hush Road, Hsinchu 30010, Taiwan}%

\begin{abstract}
We have studied how the lateral interaction affects the electric
conductance, the second-order current correlation (shot noise), and the
third-order one (skewness) of a pair of parallel Al atomic wires. The field
operator of wave function is introduced to calculate the current-current
correlations. The corresponding wave functions are self-consistently
obtained by iteration according to the Lippmann-Schwinger first-principles
calculation. The results show that when the distance between two wires is
sufficiently small, the bonding interaction near the Fermi level will induce
a spatial delocalization of electrons. This leads to an electric conduction
value greater than the sum of two uncorrelated atomic wire, and a side-band
peak is generated around the maximum conductance. This correlation can also
be observed from the shot noise and the skewness. In addition, we find that
the three-order Fano factor is negatively related to the conductance.  
\end{abstract}%


\end{frontmatter}%

\section*{Introduction}

Molecular transistors have become an interesting and popular research topic
due to recent advances in nanotechnology\cite{Ruitembeek03}. From
macroscopic aspect, the electric conductance of conductors with the same
structure is linearly proportional to the amounts of charge carriers.
However, such relationship no longer sustain because the interference
between atomic bonding and orbitals is very important when devices are
scaled down to the size of molecules. Lately, the studies on electric
conductance of molecular wires and monolayers have been performed by
mechanically controllable break junctions\cite{Ohnishi98}\cite{Yanson98}\cite%
{Hong01}and self-assembly on a solid surface\cite{Smit03}\cite{Wang10}.
Nonetheless, few have focused on the correlation of distance between the
atomic/molecular chains. As a consequence, this paper aims to investigate
the interactions between wires and the current correlations of various
moments.

According to Levitov-Lesovik transport formula\cite{Levitov93}, all moments
of carrier disturbance in the long time limit may be expanded by
transmission probability $T_{n}$(with $n$ the channel index), such as $%
S_{1}\varpropto \sum_{n}T_{n}$ and $S_{2}\varpropto \sum_{n}T_{n}\left(
1-T_{n}\right) $. The former gives the average current and the later defines
the fluctuation of current (shot noise) at zero temperature. Recently, the
third moment $S_{3}$, i.e., the asymmetry of the distribution function of
current fluctuation, is of particular interest, and has also been
experimentally probed by many groups.\cite{Gustavsson06}\cite{Bomze05}
Basically, the disturbance originates from the discontinuity of electron
itself. In Schottky's vacuum tube experiment, $S_{p}=2e\left\langle
I\right\rangle $, and $F_{2}=S_{2}/S_{p}=1$, which presents no correlation
existing between electrons. Since carriers cross the junction randomly and
independently, this electron transfer can be described by the Poisson
statistics. In nanostructures and mesoscopic devices, however, there will be
correlation between electrons in the transport process due to the electronic
structure, electromagnetic forces, the Pauli exclusion principle, and the
Coulomb interactions\cite{Jong97}. As a consequence, its Fano factor may be
greater(super-Poisson) or less(sub-Poission) than unity. The shot noise can
serve as a convenient tool for the probing of electronic structure and of
the charge amount of a unit carrier, helping understand the nature of
transport. Until now, most studies adopt model-based parameters to simulate
the transport process. Nevertheless, first-principles calculation offers a
possibility to accurately link electronic structure to all moments of the
current correlations, which is superior to the model calculations. For
instance, Chen and Di Ventra\cite{Chen03} found that the shot noise of Si
atomic wires of different lengths oscillates as a function of atom number in
the wire. Bin Wang and Jian Wang\cite{Wang10} performed first-principles
calculations based on the non-equilibrium Green's function to obtain the
shot noise properties and found that current increases almost linearly with
the bias voltage for all Pt atomic structures. All these calculated Fano
factors well coincide with the recent experimental data.

In the transport process, charge carriers enter the nanojunction from a
lead. Accordingly, it can either arrive at the second lead or reflect back.
As a prototypical example, in this paper this junction is made of a pair of
atomic aluminum wires that are sandwiched by two planar jellium electrodes.
The main reason for choosing Al atoms is because its electronic
configuration ($3S^{2}3P^{1}$) is simpler than Au atoms so that it is more
convenient to observe orbital bonding and interference of channels (three
valence electrons:\ $sp_{z}$ , $p_{x}$ and $p_{y}$\cite{Scheer97}). Besides,
studies on aluminum nanojunction have also been extensively applied to
experiments\cite{Yanson08}\cite{Zgirski08}. In the experiment by Van den
Borm and van Ruitenbeek\cite{Brom00}, the Fano factor of Al contacts was
around 0.3\symbol{126}0.6, which proves that the electrons within the
junction travel in order, instead of random motion as in classical physics.

In general, transport problems deal with non-equilibrium states, and the
counting statistics of current is often related to the bias voltage between
terminals. Therefore, these issues cannot be tackled directly by the
equilibrium ground-state of non-interaction electrodes. To overcome this
difficulty, we adopt the density functional theory (DFT) coupled the
Lippmann-Schwinger equation (LS-DFT) and perform self-consistent calculation
for the ground-state wave function of systems with nanojunctions at limited
bias voltages. This approach is essentially equivalent to the DFT combined
with the non-equilibrium Green's function\cite{Taylor01}(NEGF-DFT). One
different is that the wave function used in LS-DFT consists of the leads and
the junction, while the one for NEGF-DFT is restricted in the junction and a
self-energy has to be considered in the Green's function\cite{Wang09}. When
calculating the shot noise and skewness through the field operator of wave
function, we are faced with the products of field operators and the
Bloch-de-Dominicis theory is applied to recast these multiple operators into
pairs. The results resemble the previous study on multi-current correlation
functions by the scattering theory.\cite{Salo06}

In mesoscopic systems, the size of junction is typically greater than the
dephasing length of electrons. The wave functions of electrons will lose
coherence after multiple scattering. The interference of transverse quantum
channels is neglectably small. However, for the junction of atomic size, the
transport of electrons can be coherent. Hence the overlap integral of wave
function at different transverse momenta $\mathbf{K}$ must be kept in
describing the current correlation of different moment based on wave
functions. In other words, the effects of multi-channel mixing is included
in our calculations. Until now, the analysis of current correlations of
atom-sized parallel Al chains by first-principles hasn't been proposed. To
compele this goal, we have developed a current operator composed of DFT
self-consistent wave functions. Even though the current of higher moments is
hard to measure and calculate, it can be demonstrated to provide a way to
obtain the detailed interference effect caused by wave functions at atomic
scale. In the following, we expand the current correlation of different
moments in the basis of wave functions. The space parameters of the two
metal contacts connecting parallel Al wires are 5.8 a.u. between Al atoms,
the Al radius is $r_{s}=2$.a.u, and the last Al atom is placed $2$a.u.
outside the boundary of jellium model positive background.

\section{Theoretical Model and Calculation Method}

The Hamiltonian for the whole system is $H={H_{0}}+V\left( \mathbf{r}\right) 
$, where ${H_{0}}=\sum\limits_{E,{\mathbf{K}_{\parallel }},\alpha }{{E_{{%
\mathbf{K}_{\parallel }}}}a_{E{\mathbf{K}_{\parallel }}}^{\alpha +}a_{E{%
\mathbf{K}_{\parallel }}}^{\alpha }}$ stands for the bare electrodes
described by the jellium model and $V\left( \mathbf{r}\right) $ is the
scattering potential of the central nanostructure. The system's ground-state
wave function can be obtained through self-consistent calculation with the
Lippman-Schwinger equation, which is. 
\begin{equation}
\Psi _{E{\mathbf{K}_{\parallel }}}^{L\left( R\right) }\left( \mathbf{r}%
\right) =\Psi _{E{\mathbf{K}_{\parallel }}}^{0,L\left( R\right) }\left( 
\mathbf{r}\right) +\int {{d^{3}}{\mathbf{r}_{1}}}\int {{d^{3}}{\mathbf{r}_{2}%
}}G_{E}^{0}\left( {\mathbf{r},{\mathbf{r}_{1}}}\right) V\left( {{\mathbf{r}%
_{1}},{\mathbf{r}_{2}}}\right) \Psi _{E{\mathbf{K}_{\parallel }}}^{L\left(
R\right) }\left( {{\mathbf{r}_{2}}}\right)  \label{LSeq}
\end{equation}

Here, $\Psi _{E{\mathbf{K}_{\parallel }}}^{0,L\left( R\right) }\left( 
\mathbf{r}\right) $ represents the unperturbed wave function of the
electrodes before scattering the nanostructured object, and it can be
expressed \newline
$\Psi _{E{\mathbf{K}_{\parallel }}}^{0,L\left( R\right) }\left( \mathbf{r}%
\right) ={\left( {2\pi }\right) ^{-1}}{e^{i{\mathbf{K}_{\parallel }}\cdot 
\mathbf{R}}}\cdot u_{E{\mathbf{K}_{\parallel }}}^{L\left( R\right) }\left(
z\right) $, where $u_{E{\mathbf{K}_{\parallel }}}^{L\left( R\right) }\left(
z\right) $ is the wave propagating along the z-direction. When $z\rightarrow
\pm \infty $, the boundary conditions satisfied by wave functions on the
left and the right sides are%
\begin{eqnarray}
u_{E{\mathbf{K}_{\parallel }}}^{L}\left( z\right)  &=&\sqrt{\frac{m}{{2\pi {%
\hbar ^{2}}{k_{L}}}}}\times \left\{ {%
\begin{array}{c}
{{e^{i{k_{L}}z}}+{R_{L}}{e^{-i{k_{L}}z}},z<-\infty } \\ 
{{T_{L}}{e^{i{k_{L}}z}},z>\infty }%
\end{array}%
}\right. ,  \label{UK_L} \\
u_{E{\mathbf{K}_{\parallel }}}^{R}\left( z\right)  &=&\sqrt{\frac{m}{{2\pi {%
\hbar ^{2}}{k_{L}}}}}\times \left\{ {%
\begin{array}{c}
{{e^{-i{k_{L}}z}}+{R_{R}}{e^{i{k_{L}}z}},z>\infty } \\ 
{{T_{R}}{e^{-i{k_{L}}z}},z<-\infty }%
\end{array}%
}\right. .  \label{UK_R}
\end{eqnarray}%
\newline
The subscript ${\mathbf{K}_{\parallel }}$ is the momentum of electron with
its direction parallel to the surface of electrodes, while ${e^{i{k_{\alpha }%
}z}}$ is the free propagating wave perpendicular to it. The electron's
momentum is given by ${k_{\alpha }}=\sqrt{2m\left( {E-{\mathbf{K}_{\parallel
}}^{2}/2m}\right) -{v_{eff}}\left( {\pm \infty }\right) }$, ${v_{eff}}\left(
z\right) $ denotes the energy of the bottom conduction band in electrodes.
It should be noted that $G_{E}^{0}$ is the Green's function without the
addition of the nanostructured object. The scattering potential is given by 
\begin{equation}
V\left( {{\mathbf{r}_{\mathbf{1}}},{\mathbf{r}_{\mathbf{2}}}}\right) ={V_{ps}%
}\left( {{\mathbf{r}_{1}},{\mathbf{r}_{\mathbf{2}}}}\right) +\left\{ {\left( 
{{V_{xc}}\left[ {\rho \left( {{\mathbf{r}_{\mathbf{1}}}}\right) }\right] -{%
V_{xc}}\left[ {{\rho _{0}}\left( {{\mathbf{r}_{\mathbf{1}}}}\right) }\right] 
}\right) +\int {{d^{3}}{\mathbf{r}_{3}}{{\frac{{\delta \rho \left( {{\mathbf{%
r}_{3}}}\right) }}{{\left\vert {{\mathbf{r}_{1}}-{\mathbf{r}_{3}}}%
\right\vert }}}}}}\right\} \delta \left( {{\mathbf{r}_{1}}-{\mathbf{r}_{2}}}%
\right) .  \label{V_int}
\end{equation}%
${V_{ps}}\left( {{\mathbf{r}_{1}},{\mathbf{r}_{2}}}\right) $ is a non-local
pseudo-potential formed by covalent electrons with the core electron being
transferred. ${V_{xc}}$ is the exchange-correlation potential derived by
LDA, and the final potential is the Hatree potential. ${\rho _{0}}\left( 
\mathbf{r}\right) $ and $\rho \left( \mathbf{r}\right) $ represent the
charge density for the electrodes and the whole system, respectively. $%
\delta \rho \left( \mathbf{r}\right) $ is their difference. In numerical
calculation, the box includes the middle molecular system and partial
electrodes. It is chosen sufficiently large so that outer electrodes are not
affected by molecules. Within the box, 2880 plane waves are utilized to
calculate this coherent quantum system.

\subsection{\protect\bigskip First-principles current calculation}

At first we define a field operator, $\hat{\Psi}\left( {\mathbf{r},t}\right)
=\sum\limits_{E,{\mathbf{K}_{\parallel }},\alpha }{a_{E{\mathbf{K}%
_{\parallel }}}^{\alpha }\left( t\right) \Psi _{E{\mathbf{K}_{\parallel }}%
}^{\alpha }\left( \mathbf{r}\right) }$, to describe the many-body state of
the system. $a_{E{\mathbf{K}_{\parallel }}}^{\alpha \left( +\right) }$ is an
operator meaning the annihilation/creation of electron at time t and at
electrode $\alpha $, and it satisfies the anti-commutation relations, $%
\left\{ {a_{E{\mathbf{K}_{1}}}^{\alpha },a_{E{\mathbf{K}_{2}}}^{\beta +}}%
\right\} ={\delta _{\alpha \beta }}\delta \left( {{E_{1}}-{E_{2}}}\right)
\delta \left( {{\mathbf{K}_{1}}-{\mathbf{K}_{2}}}\right) f_{E}^{\alpha }$. $%
f_{E}^{\alpha }=1/\left\{ {1+\exp \left[ {\left( {E-{\mu _{L\left( R\right) }%
}}\right) /{k_{B}}T}\right] }\right\} $ is the probability of occupying the $%
\alpha $ electrode, which satisfies the Fermi-Dirac distribution. Based on
the definition of electric current, we derive that 
\begin{equation}
\hat{I}\left( {z,t}\right) ={{\frac{{e\hbar }}{{mi}}}}\int {d{\mathbf{r}%
_{\bot }}}\int {d{E_{1}}}\int {d{E_{2}}}\int {d{\mathbf{K}_{1}}}\int {d{%
\mathbf{K}_{2}}{e^{i\left( {{E_{1}}-{E_{2}}}\right) t/\hbar }}a_{{E_{1}}{%
\mathbf{K}_{1}}}^{\alpha +}a_{{E_{2}}{\mathbf{K}_{2}}}^{\beta }\tilde{I}_{{%
E_{1}}{\mathbf{K}_{1}},{E_{2}}{\mathbf{K}_{2}}}^{\alpha \beta }\left( 
\mathbf{r}\right) ,}  \label{CNTZ}
\end{equation}%
where $\tilde{I}_{{E_{1}}{\mathbf{K}_{1}},{E_{2}}{\mathbf{K}_{2}}}^{\alpha
\beta }\left( \mathbf{r}\right) ={\left( {\Psi _{{E_{1}\mathbf{K}_{1}}%
}^{\alpha }}\right) ^{\ast }}\nabla \Psi _{{E_{2}\mathbf{K}_{2}}}^{\beta
}-\nabla {\left( {\Psi _{{E_{1}\mathbf{K}_{1}}}^{\alpha }}\right) ^{\ast }}%
\Psi _{{E_{2}\mathbf{K}_{2}}}^{\beta }$. At zero temperature, the average
current is given by 
\begin{equation}
\left\langle {\hat{I}\left( z\right) }\right\rangle ={{\frac{{e\hbar }}{{mi}}%
}}\int {dE}\int {d{\mathbf{r}_{\bot }}}\int {d\mathbf{K}\tilde{I}_{E\mathbf{K%
},E\mathbf{K}}^{RR}\left( {{\mathbf{r}_{\bot }}}\right) ,}  \label{CNT}
\end{equation}
and it is noticed that at steady state, $\left\langle {\hat{I}\left(
z\right) }\right\rangle $ is independent of $z$.

\subsection{\protect\bigskip Second moment of the current}

\bigskip \bigskip We can extend the derivation and obtain the second and the
third moment of current. The power spectrum of shot noise has the form 
\begin{equation}
{S_{2}}\left( {\omega ,T,{z_{1}},{z_{2}}}\right) =2\pi \hbar \int {d\left( {{%
t_{1}}-{t_{2}}}\right) {e^{i\omega \left( {{t_{1}}-{t_{2}}}\right) }}%
\left\langle {\delta \hat{I}\left( {{z_{1}},{t_{1}}}\right) \delta \hat{I}%
\left( {{z_{2}},{t_{2}}}\right) }\right\rangle ,}  \label{S2def}
\end{equation}%
with $\delta \hat{I}\left( t\right) \equiv \hat{I}\left( t\right)
-\left\langle {\hat{I}}\right\rangle .$ When calculating the shot noise, we
have to deal with the quantum statistical expectation value of the products $%
\left\langle {a_{{E_{1}}}^{\alpha +}a_{{E_{2}}}^{\beta }a_{{E_{3}}}^{\gamma
+}a_{{E_{4}}}^{\delta }}\right\rangle $. With the help of Wick-Bloch-de
Dominicis theorem, it can be decomposed as $\left\langle {a_{{E_{1}}{\mathbf{%
K}_{1}}}^{\alpha +}a_{{E_{2}}{\mathbf{K}_{2}}}^{\beta }a_{{E_{3}}{\mathbf{K}%
_{3}}}^{\gamma +}a_{{E_{4}}{\mathbf{K}_{4}}}^{\delta }}\right\rangle
=\left\langle {a_{{E_{1}}{\mathbf{K}_{1}}}^{\alpha +}a_{{E_{2}}{\mathbf{K}%
_{2}}}^{\beta }}\right\rangle \left\langle {a_{{E_{1}}{\mathbf{K}_{1}}%
}^{\gamma +}a_{{E_{2}}{\mathbf{K}_{2}}}^{\delta }}\right\rangle
+\left\langle {a_{{E_{1}}{\mathbf{K}_{1}}}^{\alpha +}a_{{E_{4}}{\mathbf{K}%
_{4}}}^{\delta }}\right\rangle \left\langle {a_{{E_{2}}{\mathbf{K}_{2}}%
}^{\beta }a_{{E_{3}}{\mathbf{K}_{3}}}^{\gamma +}}\right\rangle .$
Substituting this result into Eq. (\ref{S2def}), we have%
\begin{eqnarray}
{S_{2}}\left( \omega ,T;{{z_{1}},{z_{2}}}\right) &=&{2\pi \hbar \left( {{{%
\frac{{e\hbar }}{{mi}}}}}\right) ^{2}}\sum\limits_{\alpha \beta =L,R}\int {%
dEf}_{E+\hbar \omega }^{\alpha }\left( 1-{f}_{E}^{\beta }\right) \int {d{%
\mathbf{r}_{1\bot }}}\int {d{\mathbf{r}_{2\bot }}}  \label{S2W} \\
&&\int {d{\mathbf{K}_{1}}}\int {d{\mathbf{K}_{2}}\tilde{I}_{E+\hbar \omega {%
\mathbf{K}_{1}},E{\mathbf{K}_{2}}}^{RL}\left( {{\mathbf{r}_{1}}}\right) 
\tilde{I}_{E{\mathbf{K}}_{1},E+\hbar \omega {\mathbf{K}_{2}}}^{LL}\left( {{%
\mathbf{r}_{2}}}\right) .}  \notag
\end{eqnarray}

Because shot noise reflects the current correlation at zero temperature, the
steady-state ($\omega \rightarrow 0$) shot noise is thus found as\newline

\begin{eqnarray}
{S_{2}}\left( {{z_{1}},{z_{2}}}\right)  &=&2\pi \hbar {\left( {{{\frac{{%
e\hbar }}{{mi}}}}}\right) ^{2}}\sum\limits_{\alpha ,\beta =L.R}{\int_{{E_{FL}%
}}^{{E_{FR}}}{\int {d{\mathbf{r}_{1\bot }}}\int {d{\mathbf{r}_{2\bot }}}}}
\label{S2} \\
&&\int {d{\mathbf{K}_{1}}}\int {d{\mathbf{K}_{2}}\tilde{I}_{E+\hbar \omega {%
\mathbf{K}_{1}},E{\mathbf{K}_{2}}}^{RL}\left( {{\mathbf{r}_{1}}}\right) 
\tilde{I}_{E{\mathbf{K}_{1}},E+\hbar \omega {\mathbf{K}_{2}}}^{LR}\left( {{%
\mathbf{r}_{2}}}\right) }  \notag
\end{eqnarray}

\subsection{\protect\bigskip Third moment of the current\U{ff1a}}

With similar approach, the third moment of the current is defined as 
\begin{eqnarray}
{S_{3}}\left( {\omega ,\omega ^{\prime },{z_{1}},{z_{2}},{z_{3}}}\right) 
&=&2\pi \hbar {\left( {{{\frac{{e\hbar }}{{mi}}}}}\right) ^{2}}\int {d\left( 
{{t_{1}}-{t_{3}}}\right) \int {d\left( {{t_{2}}-{t_{3}}}\right) }}
\label{S3} \\
&&\times {e^{i\omega \left( {{t_{1}}-{t_{2}}}\right) }}{e^{i\omega ^{\prime
}\left( {{t_{2}}-{t_{3}}}\right) }}\left\langle {\delta \hat{I}\left( {{z_{1}%
},{t_{1}}}\right) \delta \hat{I}\left( {{z_{2}},{t_{2}}}\right) \delta \hat{I%
}\left( {{z_{3}},{t_{3}}}\right) }\right\rangle   \notag
\end{eqnarray}%
Substituting current difference $\delta \hat{I}$ into Eq. (\ref{S3}) and
considering the steady-state current ($\omega =\omega ^{\prime }=0$) at zero
temperature ($T=0$), we have%
\begin{eqnarray}
{S_{3}}\left( {{z_{1}},{z_{2}},{z_{3}}}\right)  &=&{\left( {2\pi \hbar }%
\right) ^{2}}{\left( {{{\frac{{e\hbar }}{{mi}}}}}\right) ^{3}}\int_{{E_{FL}}%
}^{{E_{FR}}}{dE}\int {d{\mathbf{r}_{1\bot }}}\int {d{\mathbf{r}_{2\bot }}}%
\int {d{\mathbf{r}_{3\bot }}}\int {d{\mathbf{K}_{1}}}\int {d{\mathbf{K}_{2}}}%
\int {d{\mathbf{K}_{3}}}  \label{S3Z} \\
&&\times \left[ {\tilde{I}_{E{\mathbf{K}_{1}},E{\mathbf{K}_{2}}}^{RL}\left( {%
{\mathbf{r}_{1}}}\right) \tilde{I}_{E{\mathbf{K}_{2}},E{\mathbf{K}_{3}}%
}^{LL}\left( {{\mathbf{r}_{2}}}\right) \tilde{I}_{E{\mathbf{K}_{3}},E{%
\mathbf{K}_{1}}}^{LR}\left( {{\mathbf{r}_{3}}}\right) -}\right.   \notag \\
&&\left. {\tilde{I}_{E{\mathbf{K}_{1}},E{\mathbf{K}_{2}}}^{RL}\left( {{%
\mathbf{r}_{1}}}\right) \tilde{I}_{E{\mathbf{K}_{1}},E{\mathbf{K}_{2}},E{%
\mathbf{K}_{3}}}^{RR}\left( {{\mathbf{r}_{2}}}\right) \tilde{I}_{E{\mathbf{K}%
_{3}},E{\mathbf{K}_{1}}}^{LR}\left( {{\mathbf{r}_{2}}}\right) }\right]  
\notag
\end{eqnarray}%
Note that the current density in $S_{2}$ and ${S_{3}}$ keeps interference
between wave functions with different transverse momenta ${\mathbf{K}}$,
which does not coincide with Eq. (\ref{CNT}).\newline

\section{Results and Discussion}

\bigskip To analyze the conductance, the second-moment shot noise, and the
third-moment skewness, we have derived formulas for different moments of the
current correlation through the first-principles DFT combined with field
operator of wave function. That is, Eq.(\ref{CNT}), Eq.(\ref{S2}), and Eq.(%
\ref{S3Z}). The Lippmann-Schwinger equation is utilized to self-consistently
calculate the scattering wave function of the whole system. Figure \ref{dc}
shows the conductance in the unit of $G_{0}[=e^{2}/h]$ versus the distance
between two parallel wires. \FRAME{ftbpFU}{3.8614in}{2.9611in}{0pt}{\Qcb{The
differential conductance of two parallel four-atom wires as a function of
wire separation. The dashed line stands for the sum of conductance of the
two isolating atom chains. }}{\Qlb{dc}}{fig02.eps}{\special{language
"Scientific Word";type "GRAPHIC";maintain-aspect-ratio TRUE;display
"USEDEF";valid_file "F";width 3.8614in;height 2.9611in;depth
0pt;original-width 10.8102in;original-height 8.2737in;cropleft "0";croptop
"1";cropright "1";cropbottom "0";filename '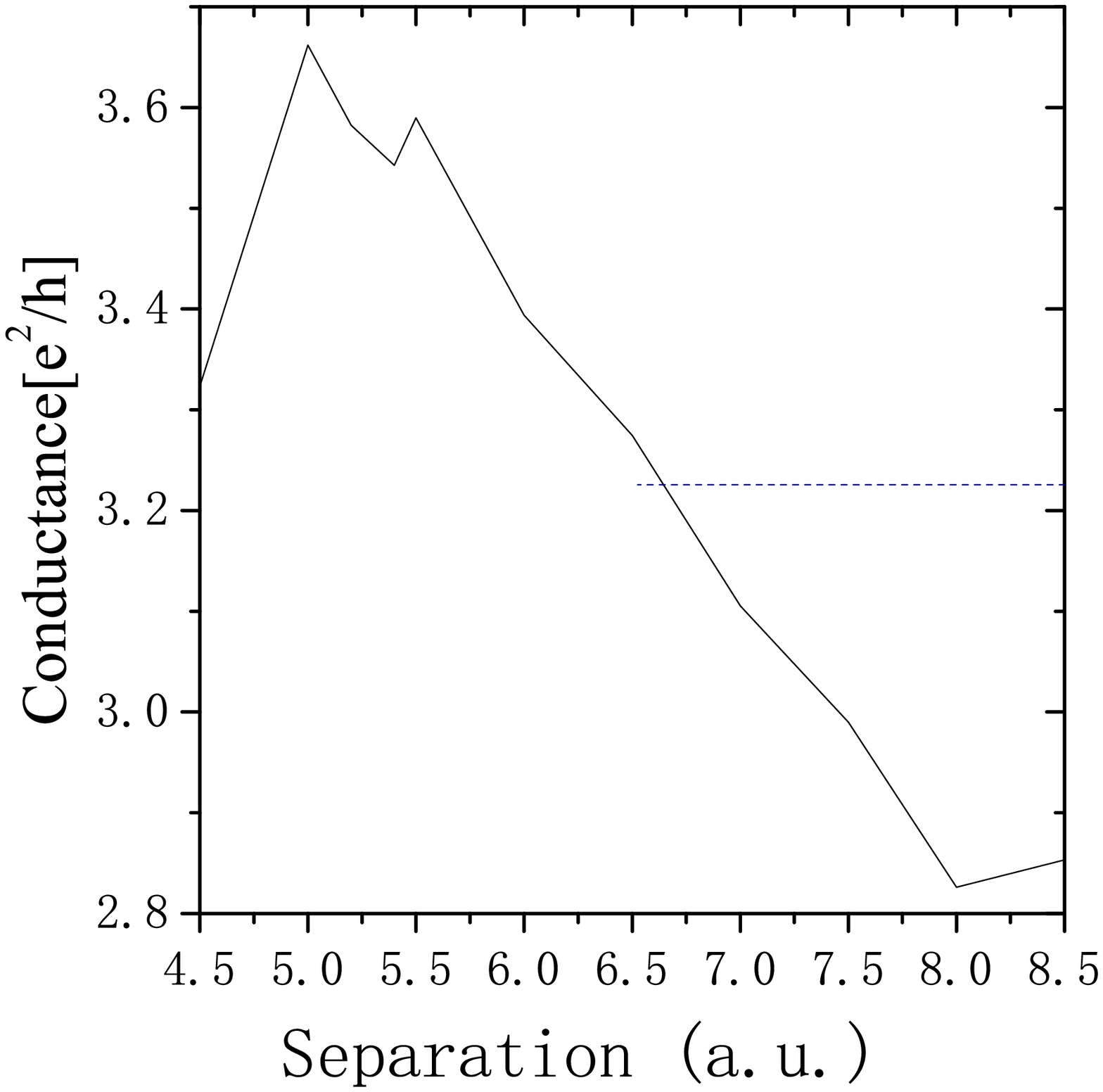';file-properties
"XNPEU";}}

We find that the differential conductance has a maximum value of 3.68 $G0$
at the distance of 5 a.u. and a side peak at the distance of 5.5 a.u. As the
distance is increased, the conductance still presents oscillating changes
(after 8 a.u). At last, the conventional conductance is calculated when the
distance between two wires is effectively infinite, i.e., two times the
conductance of a single wire. In order to investigate the modification on
the conductance by the interactions between wires, we study the distribution
of carrier density at $d=4.5\symbol{126}7.5$ a.u.

\FRAME{ftbpFU}{3.8891in}{2.9628in}{0pt}{\Qcb{The spatial distribution of
carrier density}}{\Qlb{charge}}{fig03.eps}{\special{language "Scientific
Word";type "GRAPHIC";maintain-aspect-ratio TRUE;display "USEDEF";valid_file
"F";width 3.8891in;height 2.9628in;depth 0pt;original-width
5.6948in;original-height 4.3284in;cropleft "0";croptop "1";cropright
"1";cropbottom "0";filename '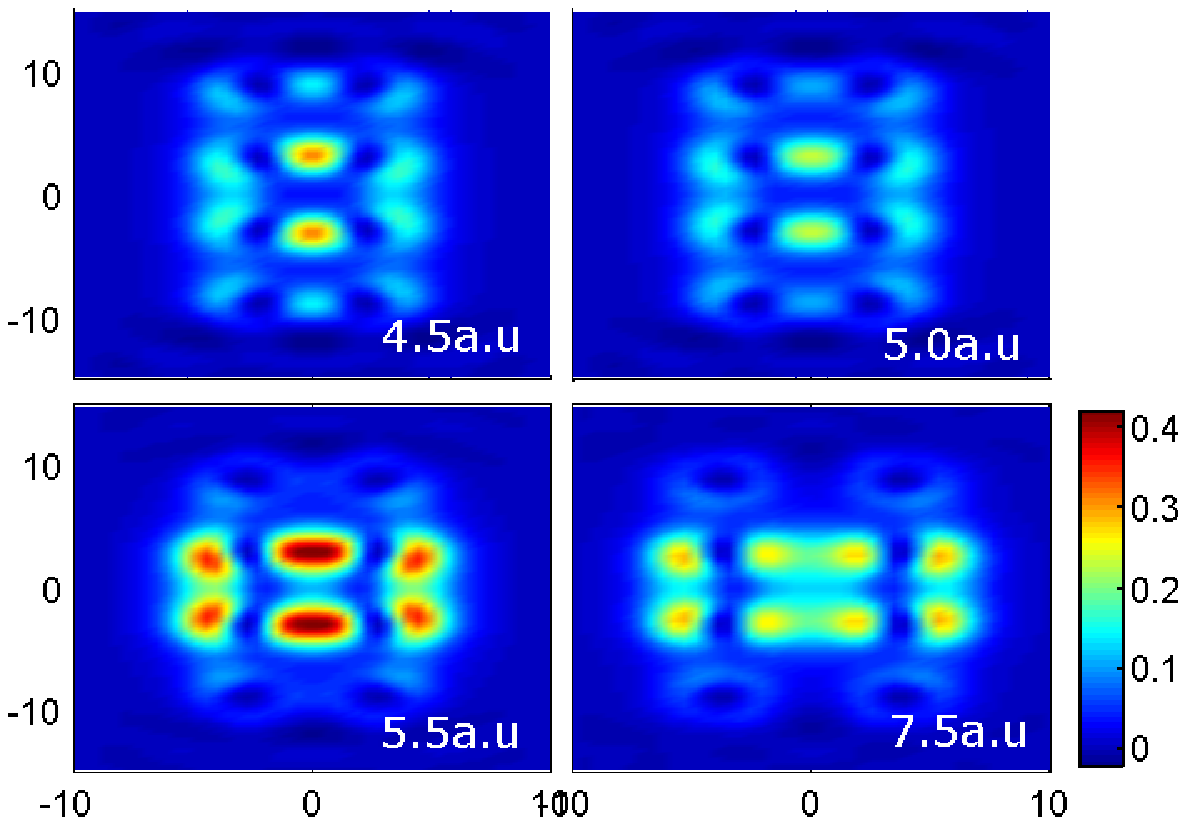';file-properties "XNPEU";}}

\bigskip From Fig. \ref{charge}, it can be seen that the $\sigma $-bonding
state is formed through the interactions between the $S$ and the $P_{z}$
orbitals in the atomic wire comprising Al atoms (3$S^{2}$3$P^{1}$), which
decides the distribution of charge carrier between wires. While the
distribution along Al wires is determined by the $\Pi $-like interaction
between the $P_{x}$ and the $P_{y}$ orbitals. A comparison of carrier
density distribution shows that the minimum density of 0.23 occurs at d=5.0
a.u. and that the minimum of 0.16 also takes place along the wire. It
indicates that the delocalization of carriers reaches its maximum at this
distance. The delocalized electrons do not belong to any bonds or atoms, and
few electrons accumulate in this region. Finally, large amounts of electrons
rapidly travel through the system and arrive at the other electrode, causing
the maximum current. We observe the distribution of DOS for further
understanding the relationship between delocalization and distance as in
Fig. \ref{DOS}. Note that a small bias voltage is exerted, , and the Fermi
energy in the left electrode is set to be zero.

Figure \ref{DOS} shows the DOS($E_{F}$) of parallel atomic chains at varied
distances. To observe its local density of states in the atomic wires, the
projected DOS is utilized. Take the case of d=$4.5$ a.u. for example, we
understand that the first two peaks (located between $E=-0.15\symbol{126}-4.5
$) are mainly contributed by the $S$ and $P_{z}$ orbitals. Nevertheless, the
orbital of $P_{xy}$ influences the region near the Fermi energy. When
increasing the distance, the bump between the first two peaks disappears
gradually and eventually become one single peak. It is noticed that the peak
shape of $P_{xy}$ orbital is modified and the extent has positive
relationship with the change in the DOS with the distance (top-left inset of
Fig. \ref{DOS}), but the peak position is not shifted. It means that the
change in carrier density caused by the distance for $\sigma $-type
interaction is far greater than that for $\Pi $-type interaction.

\FRAME{ftbpFU}{3.864in}{2.962in}{0pt}{\Qcb{The distribution of DOS in the AL
wire between two electrodes at d=$4.5,5$, and $5.5$ a.u. The top insets
present the DOS of single wires, i.e., the distance is sufficiently large.
The Fermi energy on the left side is zero, and the bias voltage is $0.01$eV.
The gray line denotes the transport regime. The top-left inset gives the DOS
at the Fermi level as a function of wire distance. The dashed bar on the
right shows the value for infinite distance. The top-left inset is the
tunneling rate of electrons near the Fermi level.}}{\Qlb{DOS}}{fig04.eps}{%
\special{language "Scientific Word";type "GRAPHIC";maintain-aspect-ratio
TRUE;display "USEDEF";valid_file "F";width 3.864in;height 2.962in;depth
0pt;original-width 10.6692in;original-height 8.1604in;cropleft "0";croptop
"1";cropright "1";cropbottom "0";filename '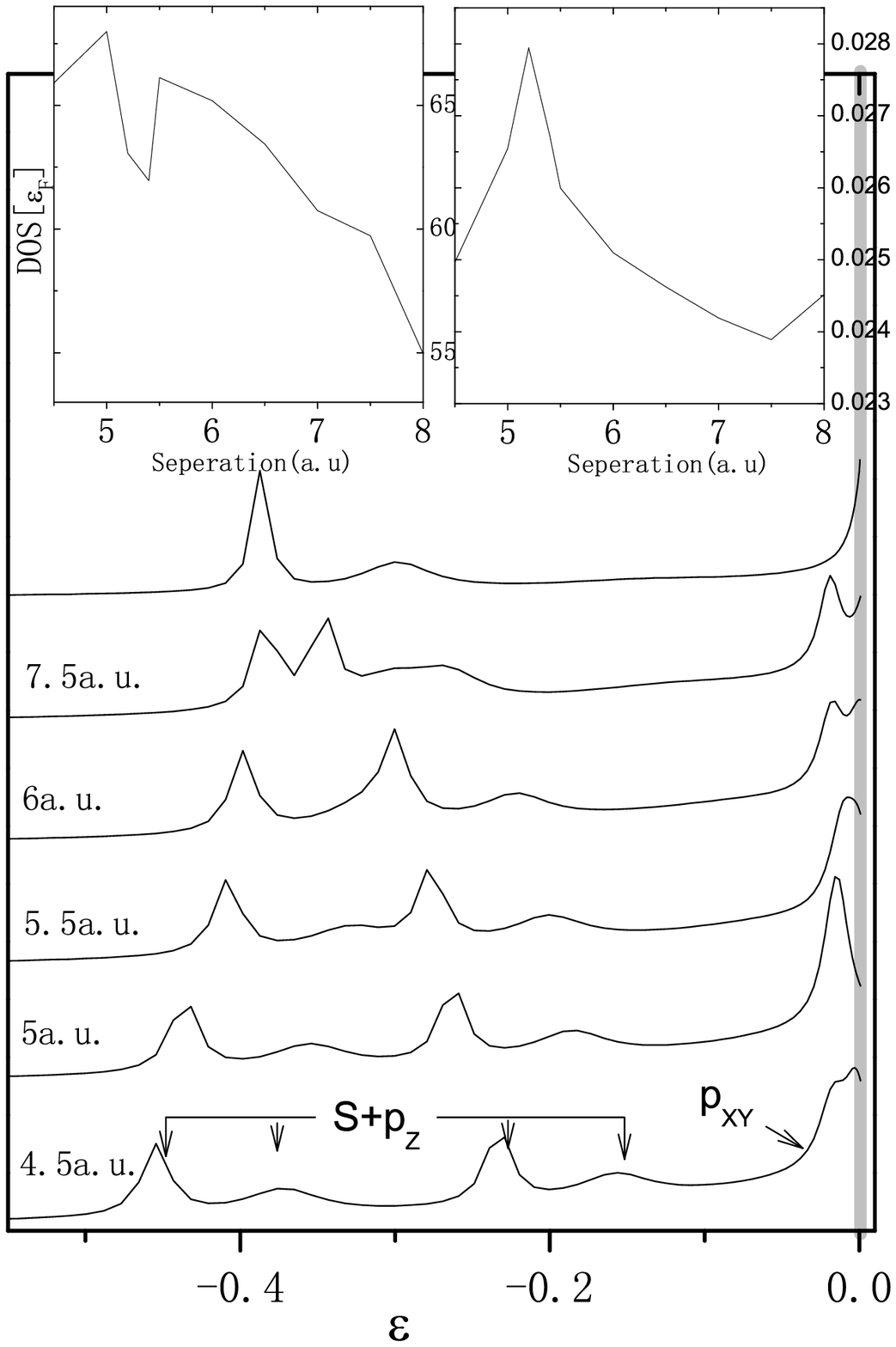';file-properties
"XNPEU";}}

Normally, the measured conductance corresponds to the change of the
electronic structure at the Fermi energy. In this case, it reflects the
exact change of $P_{xy}$ orbital. Comparing the top-left inset of Fig. \ref%
{DOS} to Fig. \ref{dc}, we find that the DOS($E_{F}$) cannot completely
decide the tendency of conductance. For example, the DOS($E_{F}$) exhibits a
parabolic curve at 5.5 to 7 a.u., forms a plateau at 7 to 7.5 a.u., and then
declines linearly. Such results disagree with those for conductance, which
shows linear decrease only. This owes to the fact that electron transport is
a dynamic process. In calculating conductance, we need to consider the
transition rate of electrons in addition to the DOS. The transition rate
versus the distance in the nanojunction is shown in the top-right inset of
Fig. \ref{DOS}. It is observed that the electronic transition rate has its
maximum at $d=5.2$ a.u., shows a valley at $7.5$ a.u., and climbs up
afterwards. So, when the DOS ($E_{F}$) is multiplied by the tunneling rate,
a side-band peak exists near the maximum delocalization distance, resulting
in a similar curve as conductance.

To realize the relationship between moments of counting statistics of
current fluctuations in the Aluminum chains, we focus on how the
second-order Fano factor (related to the shot noise) and the third-order
Fano factor (related to the asymmetry of the distribution) change with the
distance. In Fig. \ref{Fano}, $F_{2}$ first decreases and then increases,
and two dips are respectively formed at $d=5$ and $5.5$ a.u., a contrary
behavior with respect to the conductance (two peaks). 

\FRAME{ftbpFU}{3.864in}{2.962in}{0pt}{\Qcb{The second- and third-order Fano
factors versus the distance between two parallel Al chains. The right dashed
line denotes the value at infinite distance and at bias voltage of 0.01eV. }%
}{\Qlb{Fano}}{fig05.eps}{\special{language "Scientific Word";type
"GRAPHIC";maintain-aspect-ratio TRUE;display "USEDEF";valid_file "F";width
3.864in;height 2.962in;depth 0pt;original-width 10.6692in;original-height
8.1604in;cropleft "0";croptop "1";cropright "1";cropbottom "0";filename
'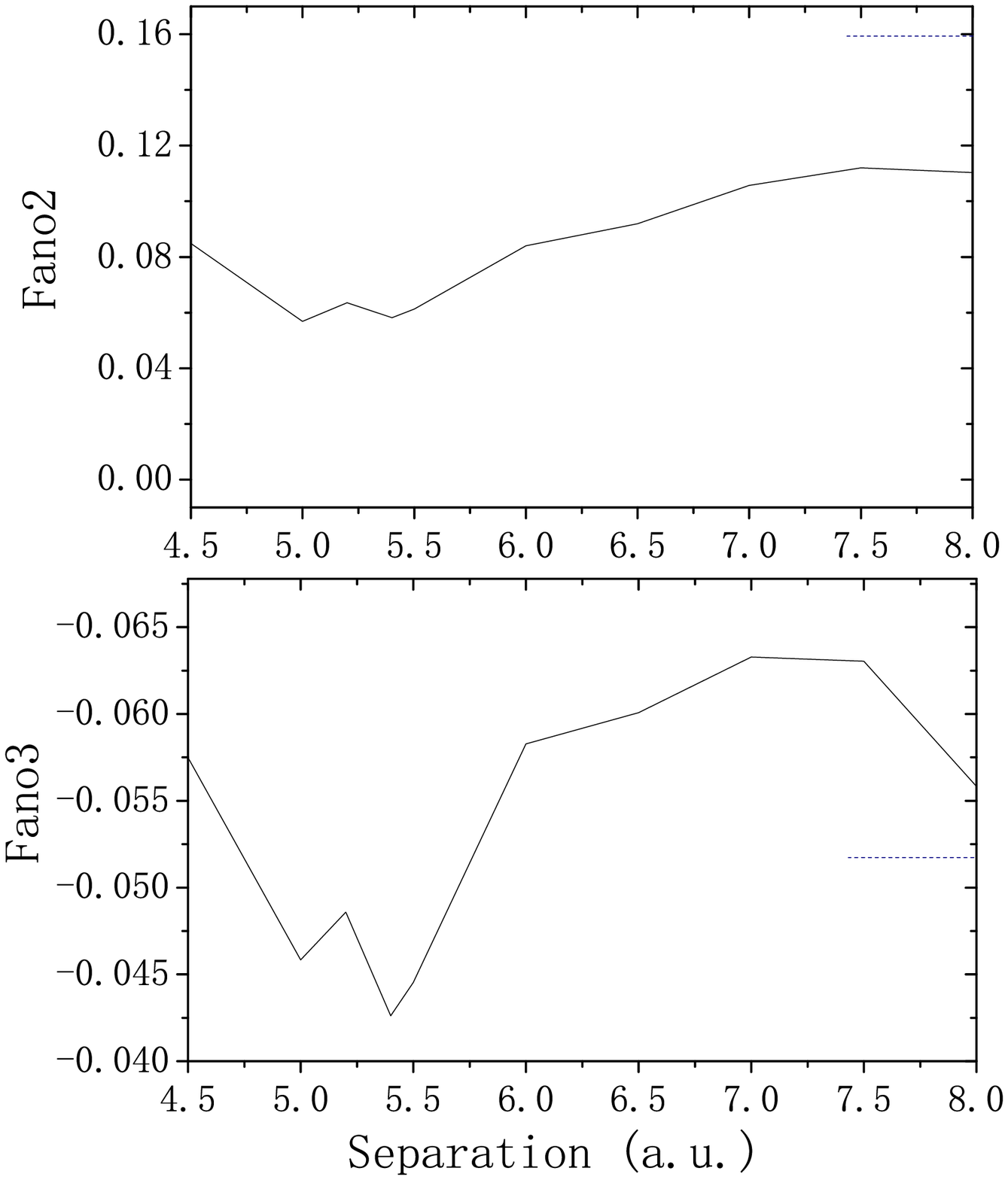';file-properties "XNPEU";}}

According to Ref. \cite{Beenakker92}, the suppression of shot noise in
metallic conductors arises from open quantum channels. Here we know that the
channels are mainly from the $Pxy$ orbital, especially at $d=5$ and $5.5$
a.u., and that the interaction between $p$ orbitals further induces a
spatial delocalization of electrons, leading to the sudden rise of current.
The bottom inset of Fig. 4 plots the third moment versus the separation
distance. It is found that $F_{3}$ is weakly negative at varied distances,
indicating the distribution function of the number of electrons transferred
through a conductor within a given time period is negatively skewed, and it
is also strongly negatively correlated with the conductance. Different from
the studies on single C4 or C5 chain, where $F_{3}$ and conductance are
always positively correlated. In this case, we find that $F_{3}$ is
positively correlated before $d=$7.5 a.u. and reveals highly negative
correlation at greater distances. It should be noticed that the least
conductance occurs at $d=$8 a.u. while the $F_{3}$ value is strongly skewed,
implying it is difficult for charge carriers to pass through the junction.%
\newline
\newline

\begin{ack}
This is the Acknowledgements section. This section is placed at the end of
the article, just before the references.
\end{ack}

\end{document}